\documentclass[pdftex,twocolumn,epjc3]{svjour3}          

\RequirePackage[T1]{fontenc}

\smartqed  

\RequirePackage{graphicx}
\RequirePackage{mathptmx}
\usepackage{amsmath}
\RequirePackage{flushend}
\RequirePackage[numbers,sort&compress]{natbib}
\RequirePackage[colorlinks,citecolor=blue,urlcolor=blue,linkcolor=blue]{hyperref}

\journalname{Eur. Phys. J. C}

\begin{document}

\title{Decay of Nambu-Goto Cosmic String Loops via Coupling to a Massive Kalb-Ramond Field
}
\subtitle{}


\author{Ivan Rybak\thanksref{e1,addr1,addr2,addr3} 
}

\thankstext{e1}{e-mail: irybak@posta.unizar.es}


\institute{CAPA \& Departamento de F\'{\i}sica Te\'{o}rica, Universidad de Zaragoza, Pedro Cerbuna, 12, Zaragoza, 50009, Spain \label{addr1}
          \and
        Centro de Astrof\'{\i}sica da Universidade do Porto, Rua
  das Estrelas, Porto, 4150-762, Portugal \label{addr2}
          \and
          Instituto de Astrof\'{\i}sica e Ci\^encias do Espa\c co, CAUP, Rua
  das Estrelas, Porto, 4150-762, Portugal \label{addr3}
}

\date{Received: date / Accepted: date}

\maketitle

\begin{abstract}
We study the dynamics of Nambu-Goto cosmic string loops coupled to a massive Kalb-Ramond field. This coupling provides a framework for describing the interaction between the cosmic string and the gauge field. Within this setup, we compute the energy flux associated with the radiation of massive Kalb-Ramond modes from oscillating string loops. For loops featuring kinks, we find that the decay time scales with the square of the loop length. In contrast, loops with cusps exhibit a decay time proportional to $3/2$ times the loop length. The results are in good
agreement with field-theoretic simulations of Nambu-Goto-like loops in
the Abelian-Higgs model, supporting the validity of the calculation.
\end{abstract}

\section{Introduction}

Soliton-like solutions emerge in various field theory models, including quantum chromodynamics, extensions of the Standard Model of particle physics, condensed matter systems, and the high-energy physics of the early universe~\cite{Tong, DauxoisPeyrard, Shnir}. In the context of cosmology, the study of one-dimensional topological solitons, known as cosmic strings, is of particular interest due to their unique properties and potential presence in various early-universe scenarios~\cite{VilenkinShellard, HindmarshKibble, JeannerotRocherSakellariadou, CopelandKibble}.

Understanding the dynamics of these non-linear objects is a difficult task. Hence, lattice field theory simulations are commonly employed as a computational tool. However, simulating cosmic strings evolving in the expanding universe presents particular difficulties, as the simulations must simultaneously capture the details of a tiny string core and the vast area of the Hubble volume. Numerous numerical approaches in field theory try to obtain realistic results in simulations that encompass such contrasting scales \cite{VincentAntunesHindmarsh,  MooreShellardMartins, HindmarshLizarragaUrrestillaDaverioKunz,   CorreiaMartins, CorreiaMartins3, DrewShellard}, mostly relying on the simplest models that correspond to global and local U(1) symmetry breaking.

An alternative simulation method approximates cosmic strings as infinitely thin and describes them using a Nambu-Goto action \cite{VanchurinOlumVilenkin,  MartinsShellard, RingevalSakellariadouBouchet, Blanco-PilladoOlumShlaer}. This approach enables more detailed simulations on a larger scale. Nevertheless, the reliability of this approximation remains uncertain as it neglects back-reaction and radiation inherent in the original field theory model \cite{B-PJ-ALL-EOUU}.

Accurate modeling of cosmic string dynamics and the evolution of their network is essential for accurate prediction of possible observational signals. Currently, gravitational lensing \cite{Vilenkin, SazhinKhlopov, SazhinaScognamiglioSazhin}, anisotropies in the cosmic microwave background \cite{LazanuShellard, LazanuShellard2, CharnockAvgoustidisCopelandMoss, LizarragaUrrestillaDaverioHindmarshKunz2, RybakSousa2}, structure formation \cite{Brandenberger:1993by, Maibach:2021qqf, Blamart:2022kly,Jiao:2023wcn} and the stochastic gravitational wave background \cite{SousaAvelino2, Blanco-PilladoOlum, CunhaRingevalBouchet, HindmarshKume, Blanco-Pillado:2024aca} serve as the most sensitive probes for detecting the presence of cosmic strings. Such knowledge allows for constraints to be imposed on high-energy models of the early universe through topological defects \cite{BattyeGarbrechtMossStoica, BattyeGarbrechtMoss, CuiLewickiMorrisseyWells, DrorHiramatsuKohriMurayamaWhite, YamadaYonekura}.

It is important to note that realistic phase transitions in the early universe may be considerably more complex than simple U(1) symmetry breaking. In particular, one may expect features such as superconductivity \cite{DavisPeter, AbeHamadaYoshioka} and Y-junctions \cite{CopelandMyersPolchinski, YamadaYonekura2} in cosmic string networks. The Nambu-Goto approximation plays a crucial role in extending the standard cosmic string network evolution to include such features~\cite{Brandenberger:1996zp, RybakAvgoustidisMartins, Rybak, CorreiaMartins2, RybakMartinsPeterShellard}, enabling the development of observational predictions beyond the simplest models~\cite{, SousaAvelino, Brandenberger:2019lfm, Cyr:2023iwu, Cyr:2023yvj, Rybak:2024djq, AuclairPeterRingevalSteer, AuclairBlasiBrdarSchmitz,Theriault:2021mrq, Rybak:2024our, Avgoustidis:2025svu}.

In this paper, we focus our attention on cosmic strings described by the Nambu-Goto action coupled to a massive Kalb-Ramond field. We organize our study as follows: In Section~\ref{Effective model of Abelian Higgs cosmic strings}, we provide the motivation for studying the Nambu-Goto action coupled to a massive Kalb-Ramond field. In particular, we construct our model as the infinitely thin limit of the Abelian Higgs cosmic string model, employing the duality between massive vector and massive Kalb-Ramond fields. This duality suggests the applicability of the model to Abelian Higgs cosmic strings. We demonstrate that the Nambu-Goto string tension can be classically renormalized due to the coupling with the massive Kalb-Ramond field, as presented in Section~\ref{Equations of motion and renormalization of tension}. Section~\ref{Radiation} discusses the emission of massive radiation from Nambu-Goto strings, focusing on specific examples such as cuspless loops~\cite{GarfinkleVachaspati} and Burden's loops~\cite{BURDEN1985277}. We compare our findings with simulations of massive radiation from oscillating Abelian Higgs cosmic string loops generated from artificial initial configurations~\cite{MatsunamiPogosianSaurabhVachaspati, HindmarshLizarragaUrioUrrestilla, PhysRevD.60.023503,Baeza-Ballesteros:2025spb}.

Throughout the paper we define a 4-dimensional metric by $g_{\mu \nu}$ with the signature $(+,-,-,-)$, where all Greek indexes run over space-time dimensions. Latin indexes in the middle of the alphabet $(i,k, \dots)$ run over spatial coordinates, while the first letters $(a, b, c, d)$ run over 2-dimensional string worldsheet coordinates.

\section{Effective model of Abelian Higgs cosmic strings} \label{Effective model of Abelian Higgs cosmic strings}

It is known that the global U(1) cosmic string model can be represented effectively via  Nambu-Goto action coupled to a massless Kalb-Ramond field \cite{VilenkinVachaspati, DavisShellard}. In this section, we demonstrate that, similarly to global strings, local U(1) Abelian Higgs strings in the infinitely thin limit can be effectively described by the Nambu-Goto action coupled to a massive Kalb-Ramond field.

\subsection{Duality between massive vector and massive Kalb-Ramond fields}

Let us consider the Abelian Higgs model, which is given by the action \cite{PhysRevLett.13.508}
\begin{equation}
\label{LagrIni}
S = \int \left( D_{\mu} \varphi \left( D^{\mu} \varphi \right)^* - \frac{1}{4} F^{\mu \nu} F_{\mu \nu} - V(\varphi) \right)  \sqrt{-g} \; d^4 x,
\end{equation}
where $V(\varphi)=\frac{\lambda}{2} \left( \eta^2 - |\varphi|^2 \right)^2$, $g$ is a metric determinant, $\lambda$ and $\eta$ define the shape of potential, $D_{\mu} = \nabla_{\mu} - i e A_{\mu}$ is the gauge covariant derivative, $A^{\mu}$ is electromagnetic 4-potential, $F_{\mu \nu} = \nabla_{\mu} A_{\nu} - \nabla_{\nu} A_{\mu}$ is electromagnetic tensor and $\nabla_{\mu}$ is a covariant derivative for curved space-time.
According to the Kibble mechanism \cite{Kibble}, the Lagrangian (\ref{LagrIni}) can give rise to topological string-like defects. 
Perturbation of the action (\ref{LagrIni}), around the minimum of potential: $ \varphi =  \eta + \frac{ \varphi_1 + i \theta }{\sqrt{2}} $, leads to
\begin{equation}
\label{LagrBroken}
S = \int \left( \mathcal{L}_{\rm br} + \mathcal{L}_{\rm int} \right) \sqrt{-g} \; d^4 x, 
\end{equation}
where $$\mathcal{L}_{\rm br} = \frac{1}{2}  \partial_{\mu} \varphi_1 \partial^{\mu} \varphi_1 - \lambda \eta^2 \varphi_1^2 - \frac{1}{4} F^{\mu \nu} F_{\mu \nu} + e^2 \eta^2 \tilde{A}_{\mu} \tilde{A}^{\mu},$$ $\mathcal{L}_{\rm int}$ includes higher order terms of $\varphi_1$ and $A_{\mu}$ 
and the vector field $\tilde{A}_{\mu} = A_{\mu} - \frac{1}{ \sqrt{2} e \eta} \partial_{\mu} \theta$ acquired mass $M = \sqrt{2} e \eta$.
By adding a divergence term to the action 
(\ref{LagrBroken}) and using equations of motion for the Proca field, one can 
obtain Hamiltonian for the vector part of the action 
(\ref{LagrBroken})\footnote{We follow the method of Ref.~\cite{DavisShellard}.}
\begin{equation}
\label{CanHam}
\mathcal{H} = - \pi^{\mu} \pi_{\mu} + M^2 \tilde{A}_0^2 - \mathcal{L}_{\rm br},
\end{equation}
where $\pi^{\mu} = F^{\mu 0}$ is a canonical momentum.
If there is no explicit time dependence in Hamiltonian, it is possible to carry out canonical transformations substituting $\mathcal{L}_{\rm br}(\tilde{A}_{\mu})$ by $\mathcal{L}_{\rm KR}(B_{\mu \nu})$ with the corresponding change of canonical coordinates $\tilde{A}_{\mu} \rightarrow B_{\mu \nu}$ and momenta $\pi^{\mu} \rightarrow \Pi^{\mu \nu} = \frac{\partial \mathcal{L}_{\rm KR}}{\partial \dot{B}_{\mu \nu}}$. To perform this canonical transformation we define\footnote{The arbitrary coefficient of transformations in Eq.~(\ref{Connection}) was chosen in a such way that the final effective Lagrangian has the massless limit coinciding with the standardly used in the literature \cite{VilenkinVachaspati, DavisShellard, BattyeShellard, BattyeShellard2, BattyeShellard3, DabholkarQuashnock, DineNicolasAkshayPatel}, but different from the one used in refs.~\cite{CopelandHawsHindmarsh, Kimyeong}.}
\begin{equation}
\label{Connection}
\tilde{A}^{\mu} \equiv \frac{1}{3! e \eta} E^{\mu \nu \lambda \rho} H_{\nu \lambda \rho}, \qquad  
F^{\mu \nu} \equiv \frac{e \eta}{ 2! } E^{\mu \nu \lambda \sigma} B_{\lambda \sigma},
\end{equation}
where $E^{\mu \nu \lambda \rho} = \frac{\varepsilon^{\mu \nu \lambda \rho}}{\sqrt{-g}}$ is a Levi-Civita tensor, and 
the field $B_{\mu \nu}$ is a massive
Kalb-Ramond field obeying equations of
motion
\begin{equation}
\label{K-REqOfMot}
\frac{1}{\sqrt{-g}} \partial_{\mu} \left( \sqrt{-g} H^{\mu \nu \lambda} \right) + B^{\nu \lambda} M^2 = 0,
\end{equation}
where $H_{\mu \nu \lambda} = \nabla_{\{\mu} B_{\nu \lambda\}} \equiv  \nabla_{\mu} B_{\nu \lambda} + \nabla_{\lambda} B_{\mu \nu} + \nabla_{\nu} B_{\lambda \mu}$.

Substituting relations (\ref{Connection})
into Eq.~(\ref{CanHam}) and subtracting
divergence term $\sim \partial_{\rho} \left( \sqrt{-g} H^{\mu \nu \rho} B_{\mu \nu} \right)$ from the Lagrangian,
we obtain the canonically transformed
Hamiltonian in terms of new variables
\begin{equation}
\label{CanHam2}
\mathcal{H} =  \Pi^{\mu \nu} \Pi_{\mu \nu} - 2 M^2 B_{0 \nu} B^{0 \nu} - \mathcal{L}_{\rm KR}(B_{\mu \nu}),
\end{equation}
where $\Pi^{\mu \nu} = H^{0 \mu \nu}$ and
\begin{equation}
\label{KRLagr}
\mathcal{L}_{\rm KR}(B_{\mu \nu}) =  \frac{1}{6} H^{\mu \nu \eta} H_{\mu \nu \eta} - \frac{1}{2} M^2 B^{\mu \nu} B_{\mu \nu}.
\end{equation}
Corresponding action (\ref{LagrBroken})
takes the form 
\begin{multline}
    \label{LagrBroken2}
S = \int \left( \frac{1}{2}  \partial_{\mu} \varphi_1 \partial^{\mu} \varphi_1 - \lambda \eta^2 \varphi_1^2  \right) \sqrt{-g} \; d^4 x + \\
\hfill \int \left( \frac{1}{6} H^{\mu \nu \lambda} H_{\mu \nu \lambda} - \frac{M^2}{2} B^{\mu \nu} B_{\mu \nu} + \mathcal{L}_{\rm int} \right) \sqrt{-g} \; d^4 x. 
\end{multline}
We showed that in the same way as for massless scalar field \cite{DavisShellard}, it is possible to carry out canonical transformation to obtain a massive Kalb-Ramond field starting from a massive vector (Proca) field \cite{SmailagicSpallucci, Barbosa}. This duality holds in a case of quantum consideration as well \cite{BuchbinderKirillovaPletnev}.

\subsection{Effective action for Abelian Higgs strings}

The action (\ref{LagrBroken2}) describes perturbed  fields around the vacuum. The complex scalar field acquires the vacuum expectation value $\varphi \rightarrow \eta$ (up to a phase factor) far away from the string core and tends to zero $\varphi \rightarrow 0$ inside the string core. Thus, the vector field $\tilde{A}^{\mu}$ (which has dual representation by Kalb-Ramond field) becomes massive outside of the cosmic string and massless inside the string core. To estimate the radius of the string we use the Compton length: $\delta_{\rm s} \propto m_{\varphi}^{-1} \propto (\sqrt{\lambda} \eta)^{-1}$, while the radius for the vector (or Kalb-Ramond) field is $\delta_{\rm v} \propto M^{-1} \propto (e \eta)^{-1}$ (details can be found in \cite{VilenkinShellard}). The relation between masses $M$ and $m_{\varphi}$ is usually defined by parameter $\beta \equiv \frac{m_{\varphi}}{M} = \frac{\sqrt{\lambda}}{e}$. 

In this section, we consider the effective description of a cosmic string. The action \eqref{LagrBroken2} can be decomposed into three parts. The first term, $S_{\rm str}$, describes the dynamics of the cosmic string, which in the limit of vanishing thickness reduces to the Nambu–Goto action. The second term, $S_{\rm 4d}$, governs the dynamics of the massive Kalb–Ramond field $B_{\mu \nu}$ and the massive scalar field $\varphi_1$ in the vacuum region outside the core of the string. The third term encodes the possible interactions between the cosmic string and the four-dimensional massive fields $B_{\mu \nu}$ and $\varphi_1$. The complete system may thus be schematically represented as
\begin{equation}
\label{LagrNamb-Got}
S_{\rm Eff} =  S_{\rm str} + S_{\rm 4d} + S_{\rm int} ,
\end{equation}
where below we define each part of the action.

The first term corresponds to a cosmic string, which can be expanded in relation to the string thickness $\delta_s$ as follows \cite{Anderson:1997ip}:
\begin{multline}
    \label{LagrNamb-Got_str}
S_{\rm str} = - \mu_0 \int \sqrt{-\gamma} \left[ 1 - \frac{\delta_s^2}{\mu_0} \alpha_1 \mathcal{R} \right. + \frac{\delta^4_s}{\mu_0} \left( \alpha_2 \mathcal{R}^2 + \right. \\
\hfil \left.  \left.  + \alpha_3 \mathcal{K}_{I \mu \nu} \mathcal{K}^{\; \mu \nu}_J \mathcal{K}^I_{\; \lambda \rho} \mathcal{K}^{J \lambda \rho}  \right)  + \mathcal{O}(\delta^6_s) \right]  d \sigma d \tau  ,
\end{multline}
where $d \sigma d \tau$ is an element of a string worldsheet, which is parametrized by $\sigma^{a} \equiv (\tau$, $\sigma)$ variables, $\gamma = \det\left[ \gamma_{ab} \right] \equiv \det\left[ X^{\mu}_{,a} X^{\nu}_{,b} g_{\mu \nu} \right]$, $X^{\mu}_{,a} \equiv \frac{\partial X^{\mu}}{\partial \sigma^a}$, $\mu_0 \sim \eta^2$ is the string tension, $\mathcal{R}$ and $\mathcal{K}_{I \mu \nu}$ are the Ricci and extrinsic curvatures of the worldsheet with $I,J$ indexes of the spatial directions orthogonal to the cosmic string. A non-trivial contribution to the string's motion occurs only at the fourth order of the expansion in relation to the string thickness, $\delta_s^4 \propto (\sqrt{\lambda} \eta)^{-4}$ \cite{Anderson:1997ip}. Therefore, we will neglect these contributions and retain only the leading order; however, see section \ref{Discussion and Conclusions } for a discussion of possible effects.

The second term consists of 4-dimensional integrals that describe the dynamics of the massive scalar field and the Kalb-Ramond field
\begin{multline}
 S_{\rm 4d} =  \int \left(   \frac{  \partial_{\mu} \varphi_1 \partial^{\mu} \varphi_1}{2} - \frac{m^2_{\varphi} \varphi_1^2}{2}+ \right. \nonumber \\
\hfill \left. + \frac{H^{\mu \nu \lambda} H_{\mu \nu \lambda}}{6}  -  \frac{M^2 B^{\mu \nu} B_{\mu \nu}}{2} \right) \sqrt{-g} d^4 x .
\end{multline}
To describe the interaction part $S_{\rm int}$, we first examine the coupling between the Kalb-Ramond field and the cosmic string. In regions distant from the core of the string, the vector field $A_{\mu}$ is characterized by a ``pure'' gauge rotation encircling the defect. This concept is illustrated, for example, in Eq.~(4.1.3) of Ref.~\cite{VilenkinShellard}. Consequently, the $4$-potential of the vector field becomes linked to the magnetic flux of the cosmic string through an integration along the path encircling the string: $\oint A_{\mu} dx^{\mu} = 2 \pi \frac{n}{e}$, where $n$ is the number of windings. Thus, to incorporate the interaction term between the cosmic string and the Kalb-Ramond field into the action defined by Eq.~(\ref{LagrNamb-Got}), one can follow a logical pathway analogous to that utilized for global strings in Ref.~\cite{DavisShellard}. 
Using relations (\ref{Connection}) and (\ref{LagrNamb-Got_str}) one concludes that equations of motion for massive Kalb-Ramond field (\ref{K-REqOfMot}) should have a source term $J^{\nu \lambda}$ of the following form \cite{BattyeShellard}
\begin{multline}
    \label{K-REqOfMot2}
\frac{1}{\sqrt{-g}} \partial_{\mu} \left( \sqrt{-g} H^{\mu \nu \lambda} \right) + B^{\nu \lambda} M^2 = -4 \pi J^{\nu \lambda} \equiv \\
\hfill \equiv-2 \pi \eta \int \left[ \delta^{(4)}(x^{\mu} - X^{\mu}(\sigma,\tau)) + \mathcal{O}(\delta_s^2) \right] d \sigma^{\nu \lambda} ,
\end{multline}
where $d \sigma^{\nu \lambda} \equiv \varepsilon^{ab} X^{\nu}_{,a} X^{\lambda}_{,b} d \sigma d \tau$ and $\mathcal{O}(\delta_s^2)$ represents terms corresponding to possible couplings between the Kalb-Ramond field and the cosmic string that arise due to the string thickness. Hence, the interaction between the Kalb-Ramond field and the cosmic string can be expressed as\footnote{Within the context of superconductivity, a comparable effective Lagrangian was proposed, as presented in Eq.~(8) of Ref.~\cite{Orland}.}
\begin{equation}
    \label{KR-NG}
S_{\rm KR-NG} = - 2 \pi \eta \int B_{\mu \nu} \left( 1 + \mathcal{O}(\delta_s^2)  \right) d \sigma^{\mu \nu}.
\end{equation}

In addition to the massive Kalb–Ramond field $B_{\mu \nu}$, there exists a massive scalar field $\varphi_1$, which can likewise be excited by the motion of the cosmic string. Massive radiation from cosmic strings has been investigated using quantum perturbative methods \cite{Srednicki:1986xg,Brandenberger:1986vj}, non-perturbative numerical simulations \cite{Olum:1999sg,HindmarshLizarragaUrrestillaDaverioKunz,DrewShellard2}, and through effective models inspired by domain wall simulations in $2+1$ dimensions \cite{B-PJ-AQU}. Although the precise form of the coupling between the Nambu–Goto action and the $\varphi_1$ field is not known, dimensional arguments suggest that the interaction must depend on the dimensionless ratio $\varphi_1/\eta$. This consideration leads to the final form of all possible interactions of the Nambu–Goto action with the massive fields:
\begin{multline}
    \label{LagrNamb-Got_KR-NG}
S_{\rm int} =  - 2 \pi \eta \int B_{\mu \nu} \left( 1 + \mathcal{O}(\delta_s^2)  \right) d \sigma^{\mu \nu} + \\
\hfill +\mu_0 \int  \left[ \mathcal{L}_{\rm \varphi-str}\left( \frac{\varphi_1}{\eta} \right) + \mathcal{O}(\delta^2_s) \right] \sqrt{-\gamma} d \sigma d \tau,
\end{multline}
where $\mathcal{L}_{\rm \varphi-str}\left( \frac{\varphi_1}{\eta} \right)$ represents the leading-order contribution to the interaction between the cosmic string and the massive scalar field $\varphi_1$, of order $\delta_s$. The interaction terms involving both the Kalb–Ramond field $B_{\mu \nu}$ and the scalar field $\varphi_1$ are of the same order, $\eta \sim \delta_s^{-1}$, in the case of critical coupling, $\beta = 1$. Although the transformations \eqref{Connection} are not valid in the massless limit $M \to 0$ $\left(\beta \to \infty \right)$, the effective action consistently reduces to the expected form for global cosmic strings in this limit.

Collecting all terms for the effective action and retaining only the leading-order terms, which are of order $\eta^2 \sim \delta_s^{-2}$ and $\eta \sim \delta_s^{-1}$, we obtain the following effective action:
\begin{equation}
\label{FullEffectAction}
S_{\rm Eff} \approx S_{\rm str} + S_{\rm int}+ S_{\rm 4d}, 
\end{equation}
where
\begin{equation}
S_{\rm str} \equiv - \mu_0 \int \sqrt{-\gamma} \, d \sigma \, d \tau ,
\end{equation}
\begin{multline}
S_{\rm int} \equiv - 2 \pi \eta \int B_{\mu \nu} \,  d \sigma^{\mu \nu} + \\
\hfill +\mu_0 \int  \mathcal{L}_{\rm \varphi-str} \left( \frac{\varphi_1}{\eta} \right) 
\sqrt{-\gamma} \, d \sigma \, d \tau,
\end{multline}

\begin{multline}
S_{\rm 4d} \equiv \int \left( \frac{  \partial_{\mu} \varphi_1 \partial^{\mu} \varphi_1}{2} - \frac{m^2_{\varphi} \varphi_1^2}{2} \right. + \\
\hfill \left. + \frac{H^{\mu \nu \lambda} H_{\mu \nu \lambda}}{6}  -  \frac{M^2 B^{\mu \nu} B_{\mu \nu}}{2} \right) \sqrt{-g} \, d^4 x. 
\end{multline}

The equation of motion for the string, up to terms of order $\mathcal{O}(\delta_s^2)$, can be written as
\begin{equation}
\label{EqOfMotString}
\partial_{a} \left( \sqrt{-\gamma} \gamma^{ab} \partial_{b} X^{\mu} \right) =  \frac{2 \pi \eta H^{\mu}_{\; \nu \rho} V^{\nu \rho} + \eta \mathcal{F}_{\varphi} }{\mu_0}   + \mathcal{O}(\delta_s^2),
\end{equation}
where $V^{\nu \rho} \equiv \varepsilon^{ab} \partial_{a} X^{\nu} \partial_{b} X^{\rho}$ and $\mathcal{F}_{\varphi}$ is a dimensionless expression that accounts for the back-reaction effect from the massive scalar field $\varphi_1$. The first term on the right-hand side represents the back-reaction due to the Kalb-Ramond field. In the case of critical coupling, when $\beta=1$, both terms on the right-hand side contribute similarly. Our main goal is to study the coupling of the Nambu-Goto cosmic string with the massive Kalb-Ramond field. In particular, in section \ref{Equations of motion and renormalization of tension}, we will examine the effect of the string tension renormalization due to massive Kalb-Ramond field. In section \ref{Radiation}, we will focus on the radiation of the massive Kalb-Ramond field, while neglecting the back-reaction effects.

\section{Green function and renormalization of tension} \label{Equations of motion and renormalization of tension}

In this section, we use Green functions to evaluate the renormalization of the Nambu-Goto string tension induced by a massive Kalb-Ramond field.

\subsection{Green functions for massive Kalb-Ramond field}

To understand the evolution of the Kalb-Ramond field, we need to study the equations of motion (\ref{K-REqOfMot2}). For this purpose, we choose Lorenz gauge conditions \cite{AuriliaTakahashi}
\begin{equation}
\label{GaugeCond}
\nabla_{\mu} B^{\mu \nu} = 0 .
\end{equation}
Thus, from (\ref{K-REqOfMot2}) one obtains
\begin{equation}
\begin{gathered}
    \label{K-REqOfMot3}
\nabla_{\mu} \nabla^{\mu} B^{\nu \lambda} + \left( g^{\nu \rho} B^{\lambda \sigma} + g^{\lambda \rho} B^{\nu \sigma} \right) R_{\sigma \rho} + \\
+ B^{\sigma \mu} \left( g^{\nu \rho} R^{\lambda}_{\sigma \mu \rho} - g^{\lambda \rho} R^{\nu}_{\sigma \mu \rho}  \right) + M^2 B^{\nu \lambda} = - 4 \pi J^{\nu \lambda} ,
\end{gathered}
\end{equation}
where $R_{\mu \nu}$ is a Ricci tensor and $R^{\lambda}_{\mu \nu \sigma}$ is a four-dimensional Riemann tensor. Hereinafter we will be considering only a local reference frame that coincides with Minkowski space, i.e. we neglect all possible corrections caused by space-time curvature. In that case, equations (\ref{K-REqOfMot3}) reduce to
\begin{equation}
\label{K-REqOfMot4}
\partial_{\mu} \partial^{\mu} B^{\nu \lambda} + M^2 B^{\nu \lambda} = - 4 \pi J^{\nu \lambda} .
\end{equation}
Solution of (\ref{K-REqOfMot4}) has the form
\begin{equation}
\label{GreenSolution}
B^{\nu \lambda}(x) = B_0^{\nu \lambda}(x) - 4 \pi \int G_M(x-x') J^{\nu \lambda}(x') d^4 x',
\end{equation}
where $B_0^{\nu \lambda}$ is a homogeneous solution and the Green function $G_M(x-x')$ is given by
\begin{equation}
\label{K-RSol3}
G_M(x-x') = -\frac{1}{(2 \pi)^4} \int \frac{\mathrm{e}^{-i k (x-x')}}{k^2-M^2} d^4 k.
\end{equation}
Introducing infinitesimal parameter we can carry out integration (see chapter IV.4 of \cite{Barut} for details):
\begin{equation}
    \label{GreenRet}
G_{M}^{r, a}(x-x') = \frac{\Theta(\pm \Delta t)}{ 2 \pi } \tilde{G}(\bigtriangleup) ,
\end{equation}
where $\tilde{G}(\bigtriangleup) \equiv \delta(\bigtriangleup^2) -  \frac{M}{2 \Delta } J_1 (M \bigtriangleup ) \Theta(\bigtriangleup^2)$, the upper sign for retarded and lower for advanced Green functions, $\bigtriangleup \equiv \sqrt{\Delta t^2 - r^2}$  ,$r = | \textbf{x} - \textbf{x}' | $, $\Delta t = t-t'$, $\mathrm{k} = |\textbf{k}|$, $\omega = \sqrt{\textbf{k}^2 + M^2}$, $\Theta(x)$ is the Heaviside step function and $J_1(x)$ is a Bessel function of the first kind. In the case of an infinitely big mass of radiation, $M \rightarrow \infty$, advanced and retarded Green functions become zero. This fact follows from the relation 
\begin{equation}
\label{GreenZero}
\lim_{M \rightarrow \infty} \frac{M}{2 x} J_1 (M x) \Theta(x^2) = \delta(x^2),
\end{equation}
which is proven in appendix A.1 of Ref.~\cite{IsoZhang}. We can obtain the self-interaction part of the Green function using the method of Refs.~\cite{Dirac, BattyeShellard3}:
\begin{equation}
\begin{gathered}
\label{GreenRet2}
 G_{\rm self} = \frac{1}{2} \left( G_{M}^r + G_{M}^a  \right).
\end{gathered}
\end{equation}
Choosing the homogeneous part of the solution $B_0^{\mu \nu}$ to satisfy boundary conditions and using the form of the source (\ref{K-REqOfMot2}) with (\ref{GreenRet2}),  we can write the self-field solution as 
\begin{equation}
\label{SolutionK-RString2} 
B^{\mu \nu}_{\rm self} = - \frac{\eta}{2 } \int  V^{\mu \nu} \tilde{G}(\bigtriangleup) d \tau d \sigma,
\end{equation}
where  $V^{\mu \nu} \equiv \varepsilon^{ab} \partial_{a} X^{\mu} \partial_{b} X^{\nu}$.

\subsection{Renormalization of the string tension}

To see how the massive Kalb-Ramond field affects the string tension, we 
should plug the expression for self-field $B^{\mu \nu}_{\rm self}$,
given by Eq.~(\ref{SolutionK-RString2}), into the action 
(\ref{LagrNamb-Got}). Considering only the last term of (\ref{LagrNamb-Got}) in the similar method as in refs.~\cite{LundRegge, DabholkarQuashnock}
we can write down 
\begin{equation}
    \label{RenormAct}
S_{\rm self} = - \pi \eta^2 \int \left( \int  \tilde{V}_{\mu \nu}  \tilde{G}(\tilde{\bigtriangleup}) d \tilde{\tau} d \tilde{\sigma}  \right) V^{\mu \nu} d \tau d \sigma, 
\end{equation}
where $\tilde{V}_{\mu \nu}$ depends on $(\tilde{\tau},\tilde{\sigma})$ and $V_{\mu \nu}$ depends on $(\tau,\sigma)$, $\tilde{\bigtriangleup}^2 = (X(\tau,\sigma) - X(\tilde{\tau},\tilde{\sigma}))^2 $. 
We consider that the cosmic string segment represented by $(\tilde{\tau}, \, \tilde{\sigma}) $  is close to the point of interest with $(\tau, \, \sigma)$, so we can perform an expansion in terms of $s = \sigma - \tilde{\sigma}$ and $t = \tau - \tilde{\tau}$ variables. In particular,
\begin{equation}
\label{Xdif}
\tilde{\bigtriangleup}^2 \approx \dot{X}^2 (\tau, \sigma) t^2 + X^{\prime \, 2}(\tau, \sigma) s^2
\end{equation}
and 
\begin{equation}
\label{VV}
V_{\mu \nu} (\tilde{\tau}, \tilde{\sigma}) V^{\mu \nu}(\tau, \sigma) \approx  2 \gamma,
\end{equation}
where we used the temporal string gauge\footnote{In the context of the specified temporal gauge (\ref{StrGauge}), we observe that the condition $\dot{X}^2+X^{\prime , 2} = 0$ may not be satisfied in the presence of a back-reaction.}
\begin{equation}
\label{StrGauge}
X^{0} = \tau, \qquad \dot{X}^{\mu} X_{\mu}^{\prime} = - \dot{\textbf{X}} \cdot \textbf{X}^{\prime} = 0,
\end{equation}
$\dot{X}^{\mu} \equiv \frac{\partial X^{\mu}}{\partial \tau}$, $X^{\prime \, \mu} \equiv \frac{\partial X^{\mu}}{\partial \sigma}$.
Using expansions (\ref{Xdif}) and (\ref{VV}), we can write down the self-interaction term in the following way
\begin{equation}
\label{RenormAct2}
S_{\rm self} = 2 \pi \eta^2 \int \sqrt{-\gamma} \Delta \mu \, d \tau d \sigma, 
\end{equation}
where $\Delta \mu = \sqrt{-\gamma} \int \tilde{G}(\tilde{\bigtriangleup}) d t d s$,
leading to a classically renormalized tension: $\mu = \mu_0 + 2 \pi \eta^2 \Delta \mu$. To obtain an explicit form of $\Delta \mu$ we use the expression
\begin{multline}
\label{DeltaIntegr}
\int \delta \left( \tilde{\bigtriangleup}^2 \right)  d t d s - \frac{M}{2 |\dot{X}|} \int \frac{J_1 (p) \Theta(p^2)}{ \sqrt{p^2 + a(s)^2}}   d p d s = \\
\hfill = \frac{1}{ \sqrt{-\gamma}} \int \frac{\mathrm{e}^{-a(s)}}{s} ds,
\end{multline}
where $p=M \sqrt{\dot{X}^2 t^2 + X^{\prime \, 2} s^2}$, $\epsilon = \sqrt{\frac{\textbf{x}^{\prime \, 2}}{1-\dot{\textbf{x}}^2}}$, $a(s) \equiv s M |\textbf{x}^{\prime}| $ and we used the property of delta function 
$\delta(f(x)) = \frac{\delta(x)}{|f'(x)|}$ for the first integral,
while for the second integral we employ the relation 
$\int_0^{\infty} \frac{J_1(x)}{\sqrt{x^2+a^2}} d x = \frac{1-\mathrm{e}^{-a}}{a}$  \cite{GradshteynRyzhik}. Hence, using the expression (\ref{DeltaIntegr}), we obtain that $\Delta \mu$ is given by 
\begin{equation}
\label{MassRenorm}
\Delta \mu 
 = \mathrm{Ei} (-M \Delta_s |\textbf{x}^{\prime}| ) -  \mathrm{Ei} (-M \delta_s |\textbf{x}^{\prime}|) ,
\end{equation}
where $\mathrm{Ei}(...)$ is the exponential  integral and we introduced two renormalization scales: $\delta_s$ - the size of a string core and $\Delta_s$ - the size of cut off, corresponding to the string length (see discussion in section II.E of Ref.~\cite{BattyeShellard3}).

There are two asymptotic cases when $M \rightarrow 0$ and $M \rightarrow \infty$, for which the tension renormalization in (\ref{MassRenorm}) reduces to
\begin{equation}
\label{MassRenorm2}
\lim_{M\rightarrow 0} \Delta \mu 
= \log\left( \frac{\Delta_s}{\delta_s} \right),  \qquad \lim_{ M, \Delta_s \rightarrow \infty} \Delta \mu \approx \beta \frac{\mathrm{e}^{-|\textbf{x}^{\prime}| / \beta }}{ |\textbf{x}^{\prime}|},
\end{equation}
were the first expression reproduces the result for global strings \cite{DabholkarQuashnock, CopelandHawsHindmarsh, BattyeShellard3}. The second expression in Eq.~(\ref{MassRenorm2}) represents the string tension renormalization for large values of $M$ and $\Delta_s$. We notice that it does not diverge when $|\textbf{x}^{\prime}| \rightarrow 0$ because of the determinant in the definition of the action (\ref{RenormAct2}). As we see from Eq.~(\ref{MassRenorm2}), the tension renormalization (\ref{MassRenorm}) is finite when $\Delta_s \rightarrow \infty$ in contrast to the global string case. Equation (\ref{MassRenorm}) illustrates that the renormalized tension undergoes variation along the string, attaining its maximum value—equivalent to the global case—at kinks and cusps.

One can try to obtain the back-reaction effect for Nambu-Goto cosmic strings similarly to the approach in refs.~\cite{CopelandHawsHindmarsh, BattyeShellard3}. However, it is impossible to write down a local equation with back-reaction corrections. As was pointed out in Ref.~\cite{IsoZhang} for the case of massive radiation from the particle, the back-reaction should be included only as integrals (\ref{SolutionK-RString2}) requiring consideration of the full system of integro-differential equations. We leave this problem for further investigation.

\section{Radiation} \label{Radiation}

In this section, we derive the equation for the radiation of massive particles from infinitely thin cosmic strings coupled to a massive Kalb-Ramond field. We also perform a comparative analysis with numerical simulations of cuspless loops \cite{MatsunamiPogosianSaurabhVachaspati, HindmarshLizarragaUrioUrrestilla} and radiation from cusps \cite{ PhysRevD.60.023503} of Abelian Higgs cosmic strings.

\subsection{Radiation for Kalb-Ramond field}

To obtain the expression for radiation of the massive Kalb-Ramond field we will follow the procedure of \cite{Nieuwenhuizen, LiRuffini, KrauseKloorFischbach}. We consider a periodically oscillating loop with the period $T_{\ell}$. Thus, we can perform Fourier transformations for the source and the field:
\begin{equation}
\label{Fourier}
J^{\mu \nu} = \sum^{\infty}_{n=-\infty} \tilde{J}_n^{\mu \nu} \mathrm{e}^{- i n \omega_0 \tau}, \qquad
B^{\mu \nu} = \sum^{\infty}_{n=-\infty} \tilde{B}_n^{\mu \nu} \mathrm{e}^{- i n \omega_0 \tau},
\end{equation}
where $\tilde{J}^{\mu \nu}_n \equiv \frac{1}{T_{\ell}} \int_0^{T_{\ell}} J^{\mu \nu} \mathrm{e}^{i n \omega_0 \tau} d \tau $ and $\omega_0 = \frac{2 \pi}{T_{\ell}}$ is a characteristic frequency.
Substituting (\ref{Fourier}) into (\ref{K-REqOfMot4}) one obtains
\begin{equation}
\label{FourierEqOfMot}
\left( \nabla^2 + k_n^2 \right) \tilde{B}_n^{\mu \nu} = 4 \pi \tilde{J}_n^{\mu \nu},
\end{equation}
where $k_n = n \omega_0 \sqrt{1-\frac{n^2_0}{n^2}} \equiv n \omega_0 \mathcal{R}_n $, $n_0 = \frac{M}{\omega_0}$. Thus, the solution of (\ref{FourierEqOfMot}) for a particular $n$-mode is given by
\begin{equation}
\label{FourierSolut}
\tilde{B}^{\mu \nu}_n = - \int G_n(\textbf{x}, \tilde{\textbf{x}}) \tilde{J}_n^{\mu \nu} (\tilde{\textbf{x}}) d^3 \tilde{\textbf{x}}, 
\end{equation}
where $G_n(\textbf{x}, \tilde{\textbf{x}}) = \frac{\mathrm{e}^{i k_n |\textbf{x} - \tilde{\textbf{x}} |}}{|\textbf{x} - \tilde{\textbf{x}} |}$.
Summing components of the field (\ref{FourierSolut}) one can write down the full expression
\begin{equation}
\label{Bsum}
B^{\mu \nu} = - \sum_{n=-\infty}^{\infty} \int \frac{\exp \left[ i n \omega_0 \left( | \textbf{x} - \tilde{\textbf{x}} | \mathcal{R}_n - t  \right)\right]}{ | \textbf{x} - \tilde{\textbf{x}} | } \tilde{J}_n^{\mu \nu} (\tilde{\textbf{x}}) d^3 \tilde{\textbf{x}}.
\end{equation}
Considering that the size of the source $\ell$, which is a loop size, much smaller than the distance $r$ to the source, we can approximate $r>>\ell$. In that case (\ref{FourierSolut}) takes the form
\begin{equation}
\label{BSumRbig}
\tilde{B}^{\mu \nu}_n = \frac{\mathrm{e}^{i k_n r}}{r} \mathrm{J}^{\mu \nu}_n + \mathcal{O}\left( \frac{1}{r^2} \right),
\end{equation}
where $\mathrm{J}^{\mu \nu}_n \equiv - \int \mathrm{e}^{-i k_n \hat{\textbf{r}} \cdot \tilde{\textbf{x}} } \tilde{J}^{\mu \nu}_n (\tilde{\textbf{x}}) d^3 \tilde{\textbf{x}}$ with a unit vector $\hat{\textbf{r}}$ oriented towards the observer.

One can notice that whenever $n_0>n$: $k_n \rightarrow i k_n$, which means an exponential decay of radiation with distance $r$. Since we assume that $r \rightarrow \infty$, we neglect all radiation coming from $n<n_0$ and higher-order terms in (\ref{BSumRbig}). Hence, substituting (\ref{BSumRbig}) into (\ref{Fourier}) and taking into account that $r \rightarrow \infty$ one obtains
\begin{equation}
\label{BSummAp}
B^{\mu \nu}_{\rm rad} = \sum_{|n|>n_0} \mathrm{J}^{\mu \nu}_n \frac{\mathrm{e}^{i (k_n r - n \omega_0 \tau)}}{r},
\end{equation}
where $\mathrm{J}^{0 i}_n =  \mathcal{R}_n \hat{r}_l \mathrm{J}^{li}_n$ due to Lorenz gauge conditions (\ref{GaugeCond}).

The Umov–Poynting vector $S \equiv T^{0i}$ for the Kalb-Ramond field in (\ref{LagrNamb-Got}) can be expressed as  
\begin{equation}
\label{UmPoynt}
\textbf{S} = 2 \left( \textbf{P} P_0 + M^2 \left( \textbf{E}_{\mathrm{v}} \times \textbf{H}_{\mathrm{v}} \right) \right),
\end{equation}
where $P^{\mu} = \frac{1}{3!}\varepsilon^{\mu \nu \lambda \eta} H_{\nu \lambda \eta}$ and three-dimensional vectors are $E_{\mathrm{v}}^{i} = B^{0 i}$, $H_{\mathrm{v}}^i = \frac{1}{2!}\varepsilon^{i k l} B_{k l}$. 
We are interested in the energy flux $I$ per solid angle $\Omega$ time-averaged over the period $T_{\ell}$, which is
\begin{equation}
\label{AverFlux}
\left\langle \frac{d I}{d \Omega} \right\rangle \equiv \frac{1}{T_{\ell}} \int_0^{T_{\ell}} \frac{d I}{d \Omega} d \tau,
\end{equation}
where $\frac{d I}{d \Omega} \equiv r^2 \hat{\textbf{r}} \cdot \textbf{S}$.
Substituting (\ref{UmPoynt}) and (\ref{BSummAp}) into (\ref{AverFlux}) and using the equality $$\frac{1}{T_{\ell}} \int_0^{T_{\ell}} \mathrm{e}^{i(k_n r - \omega_0 n \tau)} \mathrm{e}^{i(k_n' r - \omega_0 n' \tau)} d \tau = \delta_{n,-n'} $$ together with $\mathrm{J}_{-n}^{kl} = \left( \mathrm{J}_n^{kl} \right)^* $, one can obtain the final expression in the following form
\begin{multline}
\label{EnFluxRes}
\left\langle \frac{d I}{d \Omega} \right\rangle  =  \sum_{|n|>n_0} \frac{n^2 \omega_0^2}{2} \mathcal{R}_n \times \\
\left( | \varepsilon_{ikl} \hat{r}^i \mathrm{J}_{n}^{kl} |^2 + 4 \left( \frac{n_0}{n}  \right)^2 | \hat{r}_m \mathrm{J}_n^{m k} |^2 \right),
\end{multline}
which is reduced to the expression of the massless case when $M \rightarrow 0$ \cite{VilenkinVachaspati, DineNicolasAkshayPatel}.
One can notice that (\ref{EnFluxRes}) depends on the loop size $\ell = 2 T_{\ell}$, in contrast to the expressions for massless radiation. We are going to elaborate on this point in next sections for a particular shapes of loops.

\subsection{Massive gauge radiation from cuspless loops}

In this section, we conduct an analytical investigation of the massive gauge radiation from infinitely thin strings, with a specific focus on cuspless loops. Our analysis reveals that these loops, which emit only massive gauge radiation, exhibit a lifetime that is proportional to the square of the loop length: $\tau^* \propto \ell^2$. This observation aligns with findings from field theory simulations of oscillating Abelian Higgs cosmic string loops at critical coupling ($\beta = 1$), generated by artificial initial conditions.~\cite{MatsunamiPogosianSaurabhVachaspati, HindmarshLizarragaUrioUrrestilla, Baeza-Ballesteros:2025spb}. In this and the next section, we will neglect the back-reaction effects from the massive vector and scalar fields. As a result, the equation of motion for the string is given by Eq.~\eqref{EqOfMotString} with a zero right-hand side. The cuspless loop solution can then be written as \cite{GarfinkleVachaspati}:
\begin{equation}
\label{CuspLessLoop}
\textbf{X}_{\pm} = \textbf{e}_{\pm} \begin{cases} \frac{T_{\ell}}{\pi} \sigma_{\pm} - \frac{T_{\ell}}{2}, \qquad 0 \leq \sigma_{\pm} \leq \pi, \\ \frac{3 T_{\ell}}{2} - \frac{T_{\ell}}{\pi} \sigma_{\pm}, \qquad \pi \leq \sigma_{\pm} \leq 2 \pi,  \end{cases}
\end{equation}
where $\textbf{e}_{\pm}$ are unit vectors and $\sigma_{\pm} = \tau \pm \sigma$. Rewriting integrals (\ref{BSumRbig}) via $\sigma_{\pm}$ variables, we obtain
\begin{equation}
\label{In2}
\mathrm{J}^{l k}_n = \frac{\eta }{4 T_{\ell} } J_+^{[l} J_-^{k]},  
\end{equation}
where $J_{\pm}^{k} = \int_0^{2 \pi} \mathrm{e}^{i n \left( \sigma_{\pm} - \frac{\pi}{T_{\ell}} \mathcal{R}_n \hat{\textbf{r}} \cdot \textbf{X}_{\pm} \right)} d \sigma_{\pm}$.
Carrying out integration and substituting into (\ref{EnFluxRes}) we obtain the final expression  
\begin{equation}
\label{ExpressionFinal}
\left\langle  I \right\rangle \equiv \eta^2 \Gamma \equiv \eta^2 \sum^{\infty}_{n>n_0} \Gamma_n, 
\end{equation}
where $$\Gamma_n = \frac{16}{\pi^2} n^2 \mathcal{R}_n \int D^+_{n} D^-_{n} \sin^2 \alpha  \left[  \cos^2 \vartheta + \left( \frac{n_0}{n} \right)^2 \sin^2 \vartheta \right] d \Omega,$$ $\Gamma$ is the radiation efficiency, $d \Omega = \sin \vartheta d \vartheta d \phi$ is a solid angle, $D^{\pm}_{n} = \frac{1 - (-1)^n \cos \left[ \pi n \mathcal{R}_n \hat{\textbf{r}} \cdot \textbf{e}_{\pm}  \right]}{n^2 \left( 1 - (\mathcal{R}_n \hat{\textbf{r}} \cdot \textbf{e}_{\pm} )^2 \right)^2}$ and
\begin{equation}
\begin{gathered}
\label{VectBasis}
\hat{\textbf{r}} = \left\lbrace \cos \vartheta, \; \sin \vartheta \cos \phi, \; \sin \vartheta \sin \phi \right\rbrace, \\
\textbf{e}_{\pm} = \left\lbrace 0, \; \cos \alpha/2, \; \pm \sin \alpha/2 \right\rbrace.
\end{gathered}
\end{equation}
In the massless limit $M \rightarrow 0$, the expression for radiation (\ref{ExpressionFinal}) coincides with one given in Ref.~\cite{GarfinkleVachaspati}.

\begin{figure}[tbp]
\centering 
\includegraphics[width=.495\textwidth]{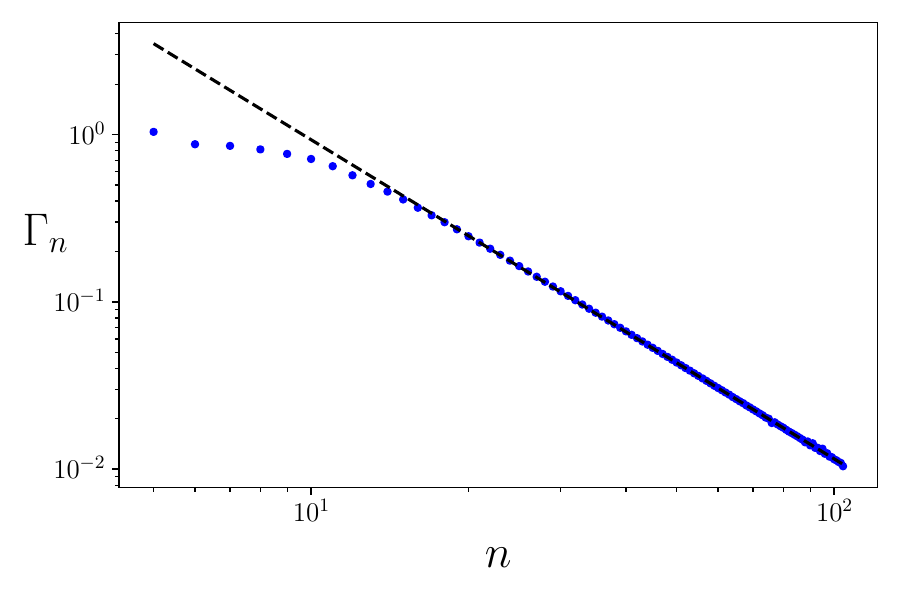}
\includegraphics[width=.495\textwidth]{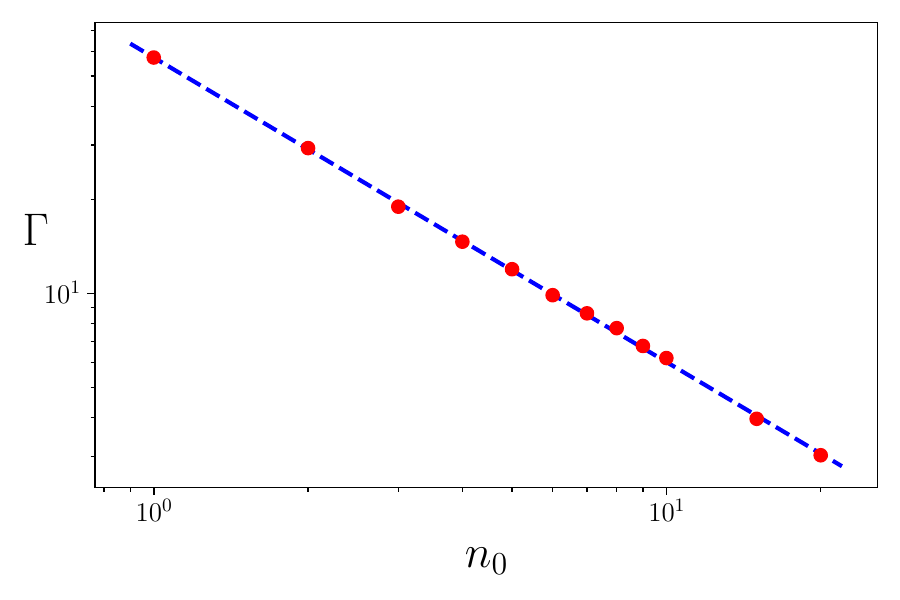}
\caption{\label{fig:1} \label{Fig1} Dots on the upper panel represent values of $\Gamma_n$ when $n_0=4$ and $\alpha=\frac{\pi}{2}$, while the dashed line is the best-fit power-law function. Dots on the lower panel represent radiation efficiency $\Gamma$ dependence on $n_0$ when $\alpha=\frac{\pi}{2}$, while the dashed line shows the best fit function $\Gamma \propto n_0^{-0.98}$. }
\end{figure}

To understand how the efficiency of radiation, denoted as $\Gamma$, depends on the loop length $\ell$, we perform the infinite sum in (\ref{ExpressionFinal}). Following the approach described in \cite{RybakSousa}, we numerically calculate and sum the first non-trivial terms of $\Gamma_n$ and approximate the remaining sum with a power-law function, as illustrated in figure \ref{fig:1}. This provides us the full radiation efficiency $\Gamma$ for various values of $n_0$, as shown in the right panel of figure \ref{fig:1}.
The best-fit value for the case of $\alpha = \frac{\pi}{2}$ is given by $\Gamma(\ell) \approx \Gamma_0  \left( \frac{M}{4 \pi} \right)^{-0.98}  \ell^{-0.98}$, where $\Gamma_0 \approx 57 $. Rounding the value, we find that the string loop decays radiating the gauge field according to
\begin{equation}
\begin{gathered}
\label{LoopDecay}
\frac{d \ell(\tau)}{d \tau} \approx - 4 \pi \Gamma_0  /  \left( \ell(\tau) M \right)  \; \rightarrow \\
\rightarrow \; \ell(\tau)^2 = \ell(0)^2 - \frac{8 \pi}{M} \Gamma_0 \tau, 
\end{gathered}
\end{equation}
which means that the lifetime of the oscillating cuspless loop, described by the Nambu-Goto action and emitting massive gauge radiation through its coupling to the Kalb–Ramond field, is given by $\tau^* = \frac{ \ell^2 M }{ 8 \pi \Gamma_0} $, i.e. the same proportionality obtained by the field-theory simulations of oscillating cuspless loops with artificially prepared initial conditions, as reported in refs.~\cite{MatsunamiPogosianSaurabhVachaspati, HindmarshLizarragaUrioUrrestilla}. It is crucial to emphasize that the number of kinks on loops influences the value of $\Gamma_0$, while the proportionality $\tau^* \propto \ell^2$ remains constant. 

In contrast, non-oscillating loops extracted from field-theoretic simulations of cosmic string networks exhibit a different scaling between the lifetime $\tau^*$ and the loop length $\ell$. As demonstrated in refs.~\cite{HindmarshLizarragaUrioUrrestilla,Baeza-Ballesteros:2025spb}, loops formed within a string network do not undergo oscillations and their lifetimes scale linearly with their length, $\tau^* \propto \ell$. In the following section, we demonstrate that the presence of cusps can further modify the relationship between $\tau^*$ and $\ell$.

\subsection{Massive gauge radiation from loops with cusps}

In this section, we explore another example of massive gauge radiation from infinitely thin strings coupled to the Kalb-Ramond field, with a specific focus on loops with cusps, particularly Burden's type of loops \cite{BURDEN1985277}. The results obtained here exhibit a decay pattern similar to that observed in numerical simulations investigating the decay of cusps \cite{ PhysRevD.60.023503}.

Loops with cusps, which satisfies equations of motion~\eqref{EqOfMotString} with a zero right-hand side due to the neglect of back-reaction effects, can be represented by the following expression \cite{BURDEN1985277}
\begin{multline}
\label{BLoops}
\textbf{X}_- = \frac{T_{\ell} }{ \pi N_-} \left( \cos N_- \sigma_-, \; \sin N_- \sigma_-, \; 0  \right), \\
\textbf{X}_+ = \frac{T_{\ell} }{ \pi N_+} \times\\
\left( \cos N_+ \sigma_+, \; \sin N_+ \sigma_+ \cos \psi, \; \sin N_+ \sigma_+ \sin \psi \right),
\end{multline}
where string worldsheet parametrization is given by $\sigma_{\pm} \in [0, \; 2\pi]$.
Substituting \eqref{BLoops} into \eqref{In2} and carrying out integrations one can obtain
\begin{multline}
\label{InegrLoop1}
J_{\pm}^{k} = 2 T_{\ell} \text{e}^{-i m \delta_{\pm}} \times \\
\begin{pmatrix} \frac{ \sin \delta_{\pm}}{d_{\pm}} J_{m_{\pm}}(m_{\pm} d_{\pm}) - i J^{\prime}_{m_{\pm}} (m_{\pm} d_{\pm}) \cos \delta_{\pm} \\ \left( \frac{\cos \delta_{\pm}}{d_{\pm}} J_{m_{\pm}} (m_{\pm} d_{\pm}) + i J^{\prime}_{m_{\pm}} (m_{\pm} d_{\pm}) \sin \delta_{\pm}  \right) \alpha_{\pm} \\  \left( \frac{\cos \delta_{\pm}}{d_{\pm}} J_{m_{\pm}} (m_{\pm} d_{\pm}) + i J^{\prime}_{m_{\pm}} (m_{\pm} d_{\pm}) \sin \delta_{\pm}  \right) \beta_{\pm}  \end{pmatrix},
\end{multline}
where $\delta_{\pm} = \arctan \left[ \frac{\hat{r}_x}{\kappa_{\pm}} \right]$, $m_{\pm} = \frac{n}{N_{\pm}}$, $d_{\pm} =  \mathcal{R}_n \kappa_{\pm} \sqrt{1 + \frac{\hat{r}_x^2}{\kappa_{\pm}^2}}$, $\kappa_- = \hat{r}_y$, $\kappa_{+} = \hat{r}_y \cos \psi + \hat{r}_z \sin \psi$, $\alpha_+=1$, $\alpha_-=\cos \psi$, $\beta_+=0$, $\beta_-=\sin \psi$. Substituting expression \eqref{InegrLoop1} into \eqref{EnFluxRes}, we numerically integrate over the solid angle $\Omega$ to obtain the expression for the radiation efficiency, $\Gamma_n$, for loops with cusps. Summing the first modes, we approximate the remaining sum with a power-law function, as illustrated in the left panel of Figure \ref{Fig2}. By performing the summation for different values of $n_0$, we demonstrate the dependence of cosmic string radiation efficiency on the string loop length, as depicted in the right panel of Figure \ref{Fig2}.

\begin{figure}[h!]
\centering 
\includegraphics[width=.495\textwidth]{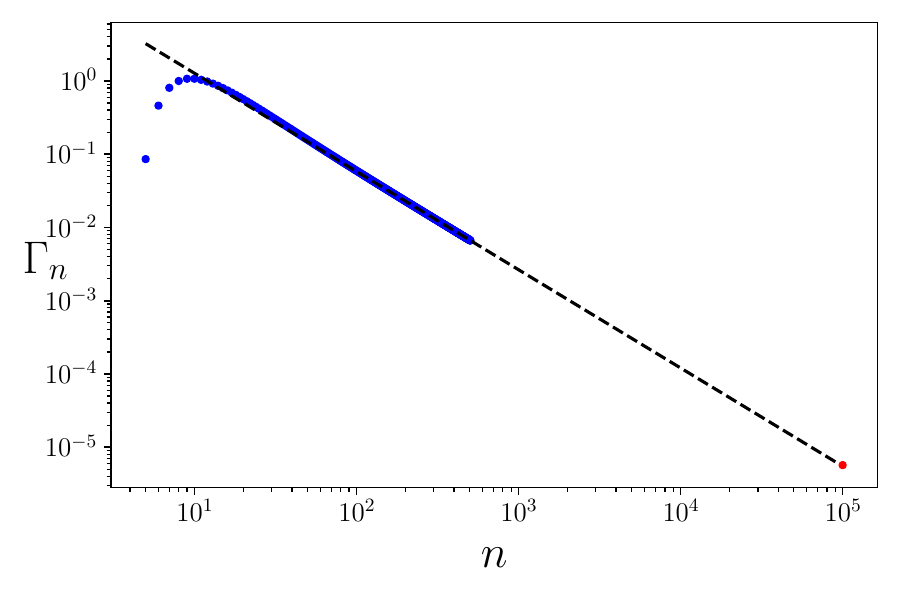}
\includegraphics[width=.495\textwidth]{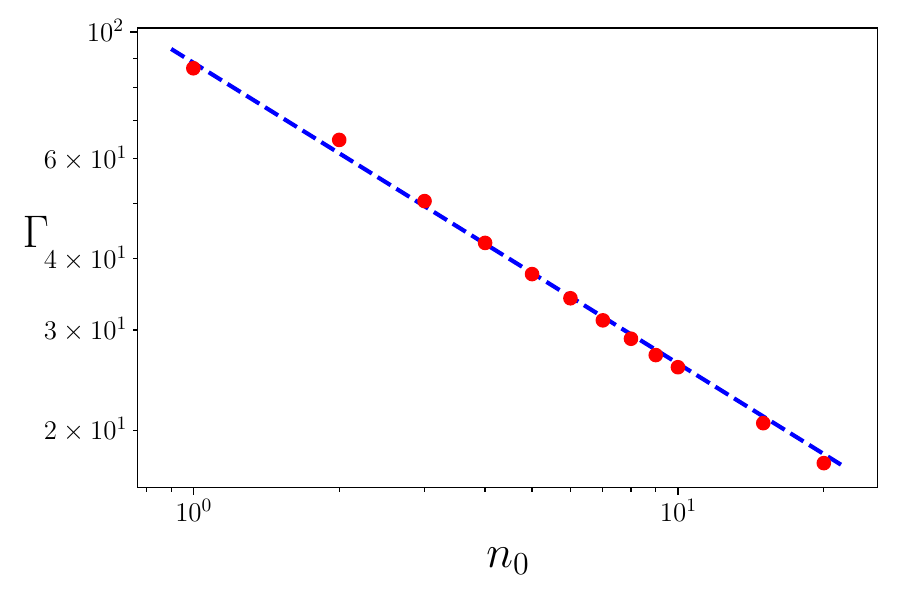}
\caption{\label{fig:2} \label{Fig2} In the upper panel, the dots represent values of $\Gamma_n$ for the case when $n_0=4$ and $\psi=\frac{\pi}{2}$, with the dashed line representing the best-fit power-law function. In the lower panel, the dots illustrate the dependence of radiation efficiency $\Gamma$ on $n_0$ with $\psi=\frac{\pi}{2}$, and the dashed line corresponds to the best-fit function $\Gamma \propto n_0^{-0.53}$.}
\end{figure}

The best-fit value for the case of $\psi = \frac{\pi}{2}$ is given by $\Gamma(\ell) \approx \Gamma_0 \left( \frac{M}{4 \pi} \right)^{-0.53} \ell^{-0.53}$, where $\Gamma_0 \approx 88 $. This radiation emission translates to the following equation describing the loop decay
\begin{multline}
\label{LoopDecay2}
\frac{d \ell(\tau)}{d \tau} \approx - \Gamma_0 \sqrt{M} / \left(  2 \sqrt{\ell(\tau) \pi}  \right)  \; \rightarrow \\
\rightarrow\; \ell(\tau)^{3/2} = \ell(0)^{3/2} - \frac{3}{4} \sqrt{\frac{M}{\pi}}  \Gamma_0 \tau.
\end{multline}
This expression leads to the lifetime of the loop emitting massive gauge radiation as $\tau^* =  \frac{4 \ell^{3/2}}{3 \Gamma_0} \sqrt{\frac{\pi}{M}} $. Interestingly, this result coincides with the one obtained by studying cusps of Abelian Higgs cosmic strings in field theory simulations \cite{ PhysRevD.60.023503}.

\section{Discussion and Conclusions } \label{Discussion and Conclusions }

In this study, we explored the dynamics of Nambu-Goto cosmic strings coupled to a massive Kalb-Ramond field. Using duality transformations, we showed that this setup effectively describes Abelian Higgs strings in the infinitely thin limit. By employing Green functions, we derived the classically renormalized tension of Nambu-Goto strings. Unlike global strings, local strings show worldsheet-dependent tension, peaking at kinks and cusps—reaching the same maximum as global strings. We also obtained integral expressions for the backreaction effects.

We then analyzed loop decay due to massive Kalb-Ramond radiation. For periodic loops, the energy flux per solid angle is given by Eq.~\eqref{EnFluxRes}, which depends on loop length $\ell$. For cuspless loops, the radiation efficiency $\Gamma$ decreases linearly with $\ell$, implying a lifetime scaling as $\tau^* \propto \ell^2$, consistent with Abelian Higgs simulations of oscillating loops generated from specific initial conditions \cite{MatsunamiPogosianSaurabhVachaspati, HindmarshLizarragaUrioUrrestilla}.
For loops with cusps, we studied burden-type solutions and found $\Gamma \propto \ell^{-1/2}$, leading to $\tau^* \propto \ell^{3/2}$, matching the simulation results \cite{PhysRevD.60.023503}.

Although our effective model reproduces simulation results, the connection to Abelian Higgs strings is non-trivial. In particular, our analysis neglects scalar modes (e.g., $\mathcal{L}_{\rm \varphi-str}$ in Eq.~\eqref{FullEffectAction}), which may contribute significantly when $\beta \sim 1$. Therefore, a complete picture should include both scalar and gauge radiation—something we leave for future work. Still, our results match simulations at $\beta = 1$, suggesting that scalar radiation either behaves similarly (as suggested by the study of massive modes for global strings \cite{Drew:2023ptp}) or is suppressed.

\begin{figure}[h!]
\centering 
\includegraphics[width=.495\textwidth]{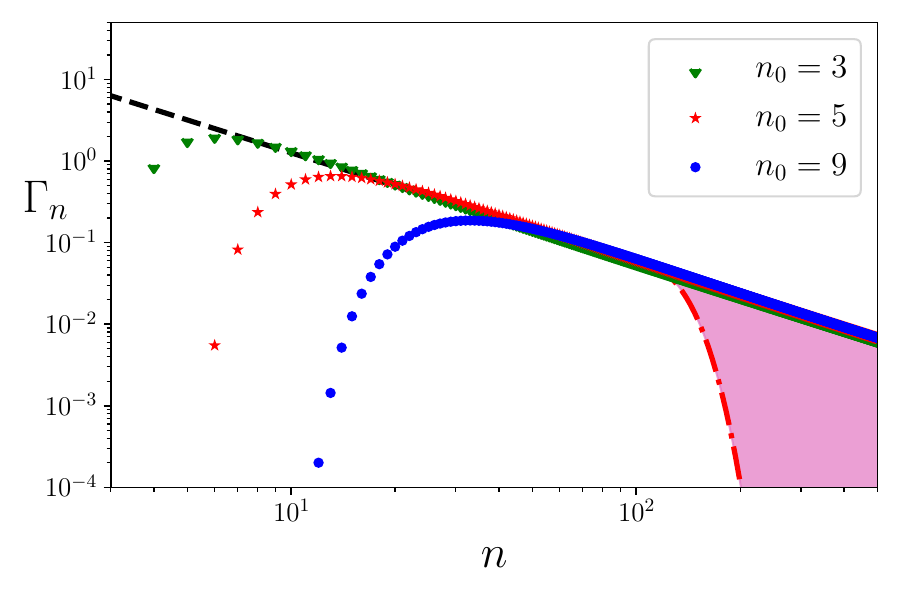}
\caption{\label{fig:3} \label{Fig2} Triangles, stars, and disks show how $\Gamma_n$ varies with $n$ for different $n_0$. The dashed line is the best-fit power-law for $\Gamma_n$ at large $n$. The dash-dotted line suggests possible exponential suppression due to string thickness.}
\end{figure}

Another deviation from the Nambu-Goto approximation arises from string thickness and back-reaction, which appear as higher-order corrections to the action \eqref{LagrNamb-Got_str} and the coupling \eqref{LagrNamb-Got_KR-NG}. These effects can smooth out kinks and cusps, leading to a cutoff at some mode $n_{\rm cut}$~\cite{OCallaghan:2010mtk}. However, if the power-law behavior of $\Gamma_n$ emerges before this cutoff, the string thickness only modifies the overall normalization $\Gamma_0$.

Importantly, Eq.~\eqref{EnFluxRes} is general and does not rely on a specific form of the source $\mathrm{J}_n^{m k}$, making it applicable to more realistic string models that account for thickness and internal modes. It relates the total radiated energy $\left\langle I \right\rangle$ to the spectral distribution $\Gamma_n$. By summing from a threshold mode $n_0$, one excludes contributions from lower harmonics. If the energy loss scales with loop length as a power law—as observed in simulations \cite{MatsunamiPogosianSaurabhVachaspati, HindmarshLizarragaUrioUrrestilla, PhysRevD.60.023503}—then $\Gamma_n$ is expected to follow a power-law behavior over the corresponding frequency range.

While our results agree with field-theoretic simulations of oscillating cosmic string loops generated from idealized initial conditions, this does not imply that such loops are produced in realistic string networks. In fact, current simulations indicate that network-formed loops generally lack sufficient angular momentum to avoid collapse and typically disintegrate before completing a single oscillation, with their lifetimes scaling linearly with length, $\tau^* \propto \ell$ \cite{HindmarshLizarragaUrioUrrestilla,Baeza-Ballesteros:2025spb}. Whether loops arising from network evolution can survive for multiple oscillations before collapsing remains an important open question \cite{RingevalSakellariadouBouchet,Blanco-PilladoOlumShlaer,Hindmarsh:2008dw}, with direct implications for predicting observational signatures, particularly gravitational waves \cite{Baeza-Ballesteros:2025spb}. By improving effective models of cosmic strings, we aim to clarify the origin of the discrepancy between field-theoretic simulations and the Nambu–Goto approximation. In particular, the backreaction and angular momentum loss included in the Nambu–Goto framework when coupled to gauge and scalar fields may provide a mechanism that suppresses the formation of oscillating loops in realistic networks.

\section*{Acknowledgements}
This work is supported by the Grant PGC2022-126078NB-C21 funded by MCIN/AEI/ 10.13039/501100011033 and ``ERDF A way of making Europe'' and Grant DGA-FSE grant 2020-E21-17R Aragon Government and the European Union - NextGenerationEU Recovery and Resilience Program on `Astrof\'{\i}sica y F\'{\i}sica de Altas Energ\'{\i}as' CEFCA-CAPA-ITAINNOVA.

We would like to express our gratitude to the organizers and participants of the Workshop ``Topological Defects in Cosmology'', held at the Jodrell Bank Centre for Astrophysics in Manchester, for their enlightening comments. We also thank Juliane F. Oliveira for her help in organizing the manuscript.

\bibliographystyle{ieeetr}
\bibliography{Biblio}

\end{document}